# Effects of Vertex Merging & Splitting on Large Coauthorship Networks: A Counterfactual Analysis

Jinseok Kim[1[0000-0001-6481-2065]]

1 University of Michigan, Ann Arbor MI 48105, USA
jinseokk@umich.edu

**Abstract.** Researchers analyze coauthorship networks, but author name ambiguity in their network data remains a significant challenge as it can change the number of vertices, distorting network properties. Although many scholars use straightforward heuristics for author name disambiguation using author's forename initials, these techniques can skew our understanding of network properties by merging or splitting vertices, raising concerns about the reliability and validity of these methods. This study investigates how different levels of vertex merging and splitting errors that are induced by name ambiguity impact network measures, using three large coauthorship networks with highly accurate algorithmic author name disambiguation. As a counterfactual scenario, two initial-based disambiguation methods widely used in coauthorship network research were applied to these datasets. Nine coauthorship network metrics were computed while varying randomly the numbers of merged or split vertices. Results show that initial-based disambiguation generates coauthorship networks with specific network properties underestimated, leading to the discovery of coauthorship networks that are smaller and more closely connected than they genuinely are. In contrast, other network metric values increase, making authors appear more collaborative and embedded within less fragmented research communities than they are. The study emphasizes the importance of careful disambiguation of vertex names in analyzing coauthorship networks for rigorous and valid findings.

## 1 Background and Importance

### 1.1 Author Name Ambiguity in Coauthorship Networks

Many researchers have conducted coauthorship network analyses using large-scale bibliographic data to model collaboration patterns across fields and nations. A persistent obstacle in such analyses is the ambiguity of author names. For example, different authors may be mistakenly grouped as one individual (a merging error), while a single author may be split into multiple distinct entities (a splitting error). Both types of errors can threaten the reliability of coauthorship network analysis.

A coauthorship network consists of vertices representing authors and edges representing their coauthorship. Authors are typically identified by the names listed in publication bylines, which creates problems when name variants (synonyms) or homonyms are involved. For example, the physicist and network scientist Mark Newman appears in publications under several variants ("Mark E. J. Newman," "M. Newman," "Mark Newman," "M. E. J. Newman"). Without disambiguation, these are treated as separate vertices, artificially splitting one individual. In contrast, when two distinct scholars share the same name, for example a physics professor and an information science professor both named 'Mark Newman' at the University of Michigan, their identities may be merged into a single vertex.

To address name ambiguity, many studies have used simple heuristics, notably *initials-based disambiguation* (IBD). IBD assumes that names with the same surname and initials represent the same author. Two variants exist: *first-initial-based disambiguation* (FIBD), which merges all names sharing a surname and first initial (e.g., "Newman, Mark E. J." and "Newman, Mark" → "Newman, M"), and *all-initials-based disambiguation* (AIBD), which distinguishes authors by all given-name initials (e.g., "Newman, MEJ" vs. "Newman, M"). Both are prone to disambiguation errors. Specifically, FIBD merges distinct authors, while AIBD splits one author into several entities. Despite these flaws, IBD has been widely used, based on the assumption that errors cancel out in large networks [1-3].

However, this assumption has been increasingly challenged. Recent studies using sampled bibliographic datasets have shown that IBD can distort the structural properties of coauthorship networks [4-6]. Despite such warnings, many published studies continue to rely on IBD. For instance, a recent survey found that about 26% of sampled coauthorship network studies still employed IBD [7]. A more serious concern is the impact on influential earlier studies that used IBD to establish widely cited findings, such as the prevalence of power-law degree distributions in collaboration networks and the principle of triadic closure (the tendency for two scholars with a common collaborator to collaborate themselves later) [1-3]. These foundational claims have rarely been revisited under alternative disambiguation methods, in part because the original datasets were not shared (e.g., the seminal paper [1] on power-law distributions of collaboration). This situation raises important concerns about the reliability and validity of coauthorship network research based on initials-based heuristics.

### 1.2 Approach and Importance of This Study

The present study seeks to raise awareness of these issues by empirically examining how different levels of name disambiguation errors affect coauthorship network measures. Specifically, we investigate how varying ratios of merged and split vertices change network properties in two large coauthorship networks.

This work integrates two lines of prior research. One line, using sampled bibliographic datasets, compared networks constructed with algorithmic disambiguation against those built using IBD [4, 5]. The other line, rooted in social network analysis, simulated how merging and splitting errors affect network properties in small social or artificial networks [8, 9]. Our study combines these approaches by simulating different levels of merging and splitting errors in empirical coauthorship networks. Importantly, to mirror the scope of earlier influential studies using IBD [1-3], we selected networks from two domains (computer science and biomedicine) covering publications from the 1990s and 2000s. While their original datasets were not available, these proxies allow us to estimate how IBD might have biased earlier findings. The results of this study are expected to help researchers make more informed choices about author name disambiguation when modeling coauthorship networks.

## 2 Study Design and Method

### 2.1 Research Design

This study aims to provide a comprehensive understanding of the potential impacts of IBD on coauthorship network properties. For this, this study compares measurements calculated on three different coauthorship networks. One network is constructed from a dataset where algorithms disambiguate author names, while the other networks are constructed from the same data but disambiguated by FIBD and AIBD. The measurements obtained from the algorithmically disambiguated data serve as a baseline and those obtained from the IBD-disambiguated data serve as counterfactuals. This approach enables us to observe how the measurements change for the same data under different disambiguation scenarios. A few social network studies have used such counterfactual analyses to test the stability and robustness of centrality measures when vertices and edges are added or removed [8, 9]. This study extends this approach to include a broader range of network measures for coauthorship networks.

### 2.2 Datasets

This study utilizes two bibliographic datasets outlined in Table 1. First, DBLP is a digital library data of conference and journal publications in computer science. The DBLP data is accessible in XML format at no cost[1]. The DBLP team disambiguates author names through algorithms and manual review [10]. The accuracy of this disambiguation process has been reported to surpass many algorithmic methods [11].

**Table 1.** Summary of Datasets

| Dataset Name | Field | No. of Papers | Publication Year |
|---|---|---|---|
| DBLP | Computer Science | 1,397,870 | 1991~2009 |
| MEDLINE | Biomedicine | 1,551,483 | 1991~2009 |

Second, MEDLINE is the PubMed data maintained by the National Library of Medicine, focusing on medical research publications[2]. For this study, papers containing the MeSH term "physiology" were selected as it is one of the most frequent MeSH terms.

[1] The dataset is available at https://dblp.uni-trier.de/xml/
[2] The raw dataset is available at https://ftp.ncbi.nlm.nih.gov/pubmed/baseline/

Author names disambiguated by algorithms were obtained from the *Author-ity* dataset[3], a version of MEDLINE that employs sophisticated machine learning algorithms and statistical models to achieve 99% accuracy in disambiguation [12, 13].

### 2.3 Measurements

**Number of Vertices.** This counts the unique author entities identified by algorithmic, FIBD, and AIBD methods.

**Number of Unique Edges.** An edge means coauthoring between vertices in a coauthorship network. Self-loops and multiple edges between two vertices are disregarded.

**Degree.** A vertex's degree indicates the number of distinct coauthors for that author.

**Density.** Network density is calculated by computing the proportion of existing edges to the total possible edges among all vertices within a network.

**Centralization.** This evaluates the concentration or variation of degree centrality values in a network, as defined in [14]. The denominator represents "the theoretically maximal sum of differences (taken pairwise between vertices)" in degree centrality.

$$C_D = \frac{\sum_{i=1}^{N}(max(D) - D_i)}{max\sum_{i=1}^{N}(max(D) - D_i)} \quad (1)$$

**Ratio of the Largest Component.** A component is a group of vertices which can connect to others through one or more connections. The ratio of the number of vertices in the largest component to the total number of vertices in the network is the ratio of the largest component.

**Average Shortest Path Length.** The shortest path length (or geodesic) is the minimum number of edges connecting two reachable vertices in a component. Calculating the average shortest path length in large coauthorship networks is computationally expensive, with time complexity increasing by $O(|V|+|E|)$, where $V$ is the number of vertices and $E$ the number of edges. To address this, we approximate geodesics using the robust method in [4], which estimates the average by computing shortest paths from a random sample of 1,000 vertices to all others in the network.

**Degree Assortativity.** This evaluates how authors collaborate with others with similar degree centrality values. In technical terms, it is calculated as the Pearson correlation coefficient of the degrees at both ends of an edge between vertex pairs [15].

**Clustering Coefficient.** : This metric, a.k.a. transitivity, represents the ratio of triadic closure among vertices. In coauthorship networks, this indicates the likelihood of an edge forming between two vertices that share a common neighbor [2].

---

[3] The dataset is available at https://databank.illinois.edu/datasets/IDB-2273402

$$CC = 3 \times \frac{number\ of\ triangles\ on\ the\ network}{number\ of\ connected\ triples\ of\ vertices} \quad (2)$$

### 2.4 Error Simulation

Author names in each dataset are disambiguated using algorithms, FIBD, and AIBD. Algorithmically disambiguated names obtained from *Author-ity* (MEDLINE) and DBLP public release are the proxy of the gold standard. FIBD and AIBD are implemented by converting an author name string into the format of a full surname followed by the first forename initial (FIBD) and a full surname followed by all available forename initials (AIBD). Table 2 reports the disambiguation errors by IBD for each dataset when all names are disambiguated. In DBLP, for instance, 817,628 distinct authors were recognized by algorithmic disambiguation. About 373,000 authors (46%) were unaffected by IBD, while 444,597 entities (54.38%) were merged by it.

**Table 2.** Summary of Disambiguation Errors

| Disambiguation Method | Total Entities (%) | No Errors (%) | Merging (%) | Splitting (%) | Merging & Splitting (%) |
|---|---|---|---|---|---|
| **DBLP-FIBD** | 817,628 (100) | 373,031 (45.62) | 444,597 (54.38) | - | - |
| **DBLP-AIBD** | | 486,250 (59.47) | 331,378 (40.53) | - | - |
| **MEDLINE-FIBD** | 1,839,407 (100) | 722,145 (39.26) | 1,073,772 (58.38) | 18,018 (0.98) | 25,472 (1.38) |
| **MEDLINE-AIBD** | | 922,523 (50.15) | 752,651 (40.92) | 82,458 (4.48) | 81,775 (4.45) |

To evaluate the effect of disambiguation errors by IDB on coauthorship networks, this study introduces progressively disambiguation errors into a coauthorship network, adopting the simulation ideas from [4, 16]. The process involves the following steps:

*Step 1.* A list of author entities from algorithmically disambiguated data was acquired.

*Step 2.* Each author entity on the list received a name ambiguity label for merging, splitting, or merging & splitting by FIBD and AIBD. The labels are mutually exclusive: Each IBD assigns only one label to an author entity.

*Step 3.* Author entities were randomly chosen from the list, resulting in *N*% of entities having a specific ambiguity label.

*Step 4.* All name instances linked to the randomly selected author entities were altered to first-initial or all-initial formats.

*Step 5.* Network metrics were calculated for coauthorship networks created from algorithmically disambiguated data and the same data disambiguated by IBD.

*Step 6.* Steps (1) through (5) were repeated for *N*, ranging from 0% to 100% in 1% increments.

### 2.5 Measurement Error Ratio

Measurement errors are quantified by calculating a property change ratio which is defined as follows.

$$ErrorRatio\ (\%) = \frac{Val_A - Val_B}{Val_B} \times 100 \tag{3}$$

Here, $Val_A$ refers to the value of a network property calculated on the coauthorship network disambiguated by IBD. In contrast, $Val_B$ refers to the value by algorithmic disambiguation. This study considers disambiguation errors acceptable if they result in less than a 5% error in any network measure. This threshold is arbitrary and intended for illustration purposes.

## 3 Analysis

### 3.1 Merging Effects

Figures 1 and 2 display the alterations in nine metrics when *N*% of merging occurs in three data using FIBD (green triangles) and AIBD (red crosses). An inset image in each subfigure demonstrates the ratios of merging that result in 5% or lower measurement errors. The highest ratio of merging that yields 5% or less measurement error is marked in green for FIBD or red for AIBD. In each figure, the *y*-axis values at zero percent on the *x*-axis represent the network property values calculated on the coauthorship network constructed from algorithmically disambiguated data. These values serve as anchor points for comparison with any changes induced by IBD.

All figures reveal a few common trends. First, as the proportion of merged author entities rises from zero to 100%, the measurement values of specific network metrics, such as average degree, centralization, density, and the ratio of the largest component, increase (i.e., plots shift upward and right). This indicates that these network properties are overestimated for coauthorship networks when relying on IBD.

Second, other metrics like the count of vertices, the count of edges, transitivity, and average shortest path lengths decrease as the proportion of merged entities increases (i.e., plots shift downward and right). This implies that IBD leads to an underestimation of these network properties. These tendencies, i.e., over or underestimating the mentioned network properties due to merging, are consistent with findings from previous studies for generic social networks and coauthorship networks [4, 8, 9].

Third, the changes induced by FIBD are more significant than AIBD across most network metrics. This is because FIBD generates more merged authors than AIBD. This suggests that, for the same *N*% entity merging, FIBD creates more merged author entities than AIBD, resulting in more alterations of network properties. For instance, centralization continues to grow as the proportion of merged entities increases since merged vertices typically have larger degrees and contribute to an increased concentration of degrees in the coauthorship networks. As FIBD merges more entities, the centralization values with FIBD increase more than those with AIBD.

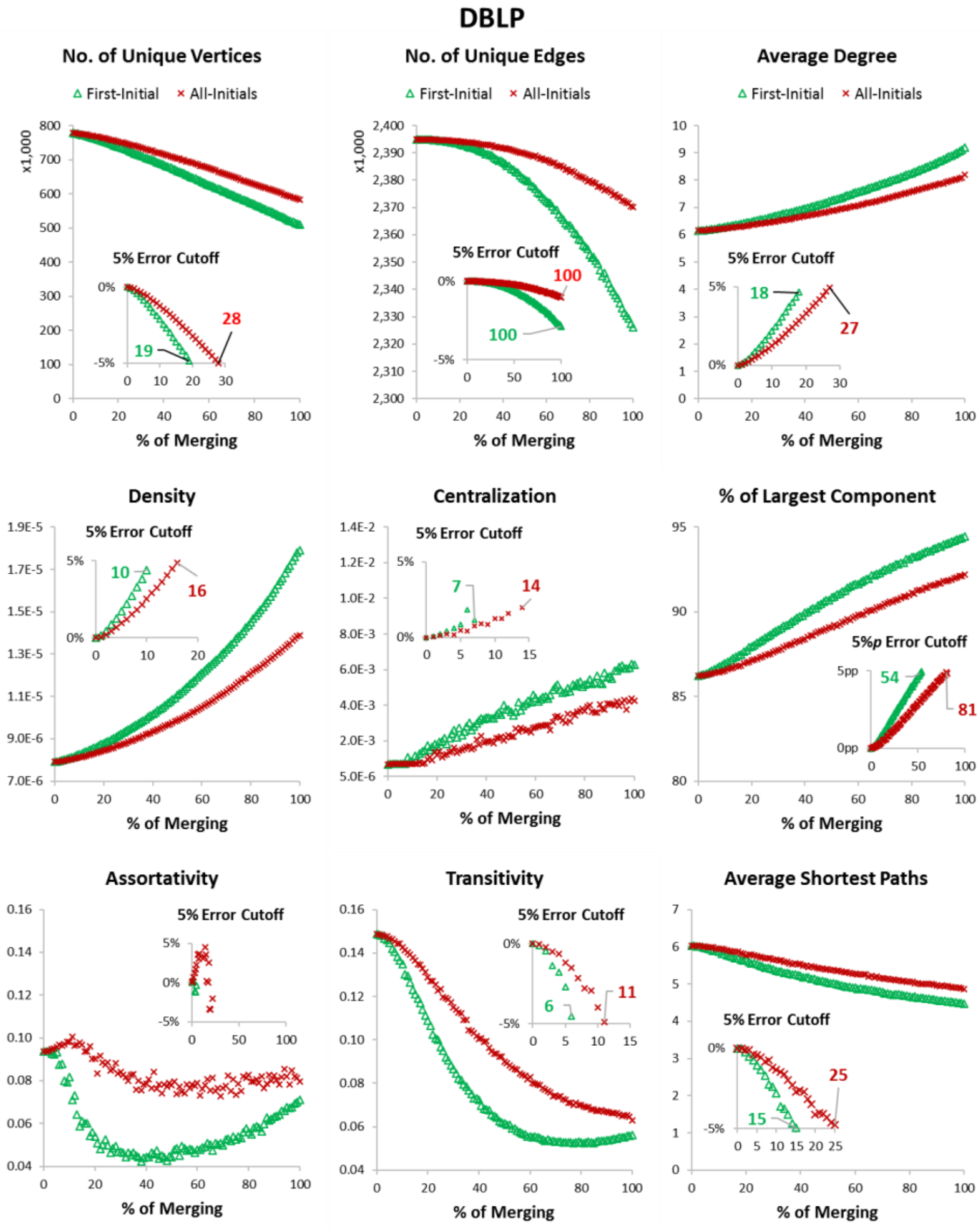

**Fig. 1.** Changes of Coauthorship Network Properties at Various Merging Levels for DBLP

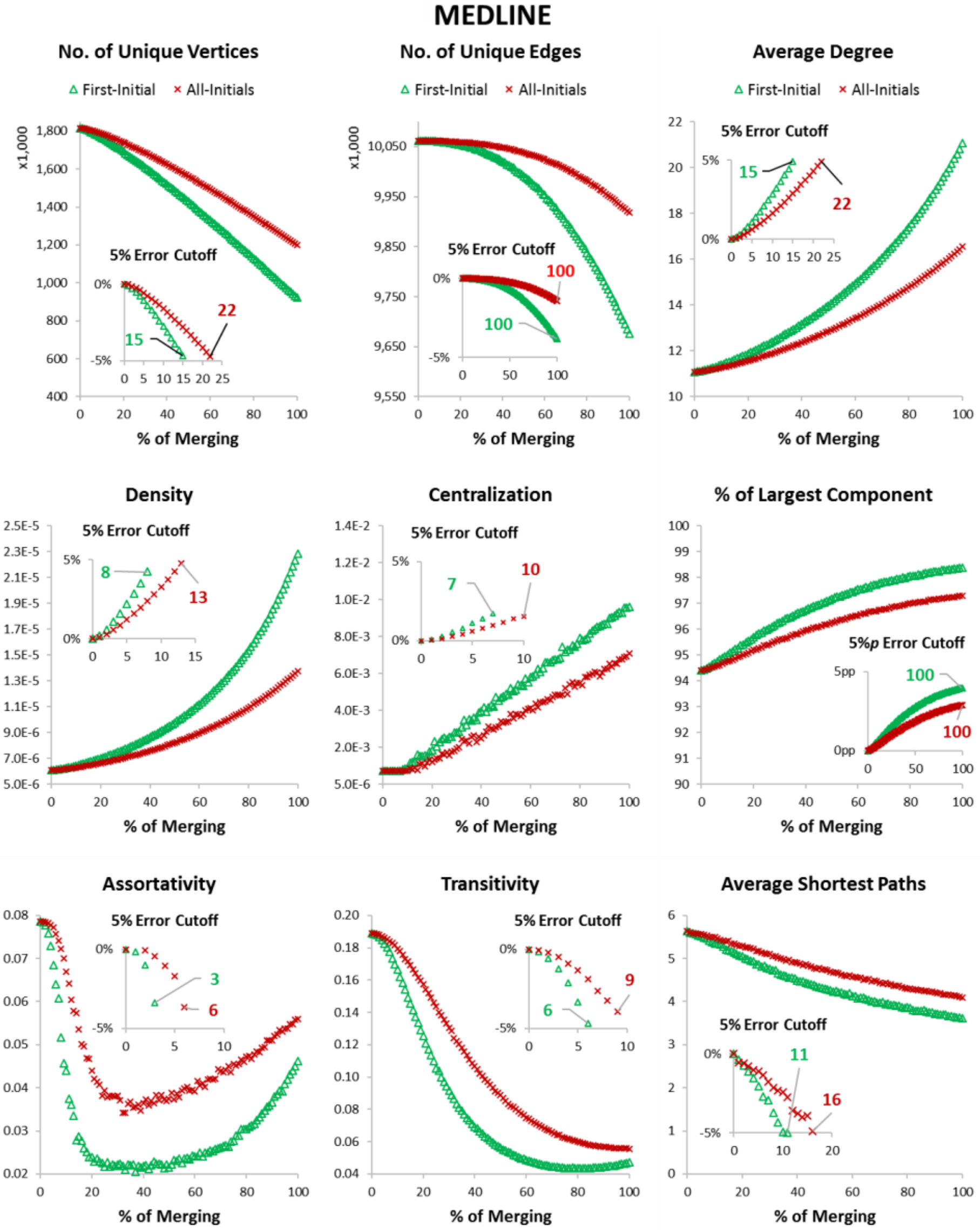

**Fig. 2.** Changes of Coauthorship Network Properties at Various Merging Levels for MEDLINE

A few instances warrant more in-depth analysis. First, the vertex count changes nearly monotonically with the ratio change, while the edge count displays a descending curve as the proportion of merged entities increases. This can be attributed to the consolidation of edges between merged authors and their coauthors who share similar first or middle name initials.

Second, degree assortativity shows a U-shaped pattern. This study calculated degree assortativity using Pearson's correlation coefficient. This method posits that assortativity increases when vertices with similar degrees are more likely to be connected. However, the computation is recognized for its sensitivity to vertices with high degrees. Even if most vertices in a network have neighbors with similar degrees, outliers with high degrees and numerous low-degree neighbors can deflate the overall assortativity. So, the U-shaped variation in assortativity can be ascribed to alterations in degree similarity among vertices with high degrees because of heightened merging. In specific merging ratios, smaller-degree vertices merged to form high-degree vertices with low-degree neighbors, reducing the overall assortativity. As merging intensified, high-degree vertices began to connect, increasing assortativity levels.

Concerning the acceptable measurement inaccuracies (5% threshold), some metrics were more susceptible to merging than others. For instance, the edge count exhibited less than 5% errors even when all author entities were vulnerable to merging by IBD in both datasets (refer to the inset figures in the top middle subgraphs of Figures 1 and 2). In contrast to the edge count, only 6-11% of IBD-induced merging resulted in a 5% error rate in transitivity for both datasets. This suggests that the impact of merging errors can be assessed differently depending on the network measures.

### 3.2 Splitting Effects

For splitting, author entities were randomly selected to simulate errors from 0 to 100 percent. This approach was used because in practice, merging dominates while splitting is minimal. For example, if a unique author has ten name instances affected by IBD, typically nine are merged and only one is split. Including such cases in simulations would cause merging effects to overshadow the impact of splitting.

Also note that, as shown in Table 2, splitting did not occur in DBLP, not because its author names are immune to it but due to the structure of DBLP data. DBLP replaces original name strings with a single disambiguated form for all an author's publications. In addition, it assigns each unique author a standardized name string, sometimes with a four-digit suffix to distinguish homonyms (e.g., Wang Wei 0001, Wang Wei 0002). Thus, a unique author cannot appear under multiple name strings in the public release. So, IBD-induced splitting did not arise in our study.

Figure 3 presents the variations in coauthorship network properties for MEDLINE when *N*% of author entities are split. Unlike merging, the effect of splitting on coauthorship networks is more pronounced for AIBD (red crosses) compared to FIBD (blue triangles). This is anticipated as AIBD is more prone to splitting than FIBD.

Moreover, in contrast to the significant changes in network properties caused by merging, splitting exerts a minor influence on these properties. With 100% splitting, for example, many metrics resulted in 5% or less measurement errors for MEDLINE. For instance, when disambiguated by FIBD, a 100% splitting scenario resulted in measurement errors of less than 5% across nine network metrics. With AIBD, six metrics (the vertex count, the edge count, average degree, centralization, transitivity, and the ratio of the largest component) exhibit measurement errors of less than 5%. This suggests that relative to the errors resulting from merging as shown in Figures 1 and 2, measurement errors caused by IBD were less susceptible to the effects of splitting compared to merging. This finding aligns with the findings observed in [4, 16].

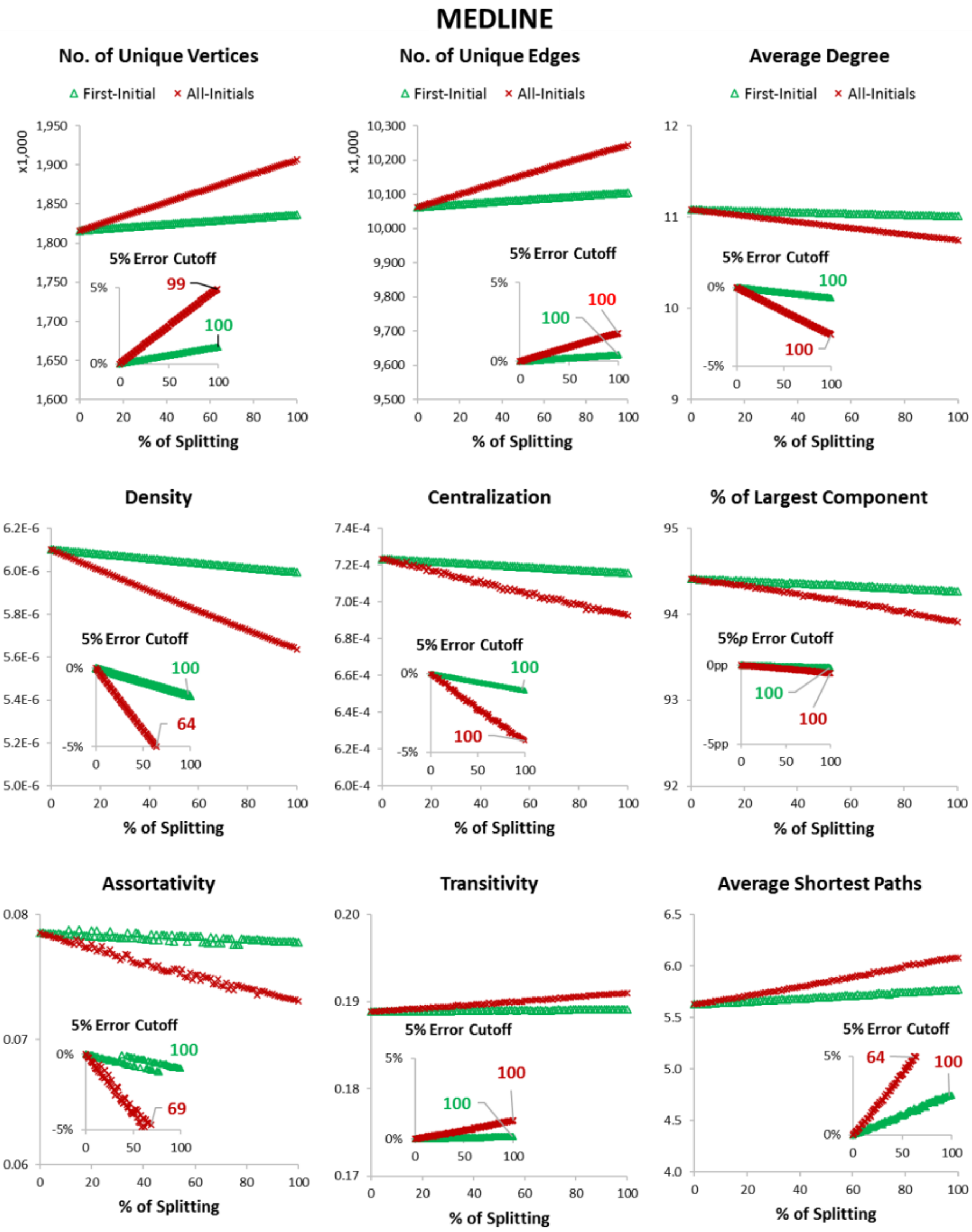

**Fig. 3.** Changes of Coauthorship Network Properties at Various Splitting Levels for MEDLINE

More importantly, the alterations in most network measures due to IBD were almost linear. This suggests that the error magnitude caused by splitting can be estimated with decent accuracy for a given level of splitting. This situation, however, does not hold for merging. Although it is feasible to determine the merging levels that meet a particular error rate (in this case, 5%) via simulation, the curvature of the merging simulation plots indicates that it is difficult to determine the error rate by relying on a particular merging level.

## 4 Conclusion and Discussion

This study shows how author name disambiguation choices affect our understanding of coauthorship network properties. Two datasets, DBLP and *Author-ity* (MEDLINE), were used, each with highly accurate algorithmic disambiguation. To create counterfactual scenarios, two initials-based methods, first-initial (FIBD) and all-initials (AIBD), were applied, reflecting practices in coauthorship network research.

Nine metrics were calculated for both algorithmically disambiguated networks and those using FIBD and AIBD, while varying the proportions of merged and split entities. Compared to algorithmically disambiguated networks, IBD-based networks underestimated the number of vertices, assortativity, transitivity, and average shortest path length, suggesting smaller, denser, and less collaborative networks. Meanwhile, they overestimated average degree, density, centralization, and the size of the largest component, making authors appear more productive and integrated than they actually are. The magnitude of distortion grew with the extent of merging and splitting.

These findings align with earlier studies overall but extend prior work by testing large empirical datasets across two domains and simulating varying error levels. Results highlight that splitting distorts network properties less than merging, implying that stricter rules in heuristic disambiguation or conservative parameter tuning in machine learning methods are preferable. When uncertain, it is safer to treat name instances as distinct rather than incorrectly merging them.

This study cautions researchers to carefully consider disambiguation methods when analyzing coauthorship networks. While many scholars have moved toward algorithmic disambiguation [7], IBD remains common. Algorithmic methods more closely approximate ground truth, but they require computational expertise, feature selection, and implementation skills that may not be always available for coauthorship network researchers. This underscores the need for accessible tools that enable researchers without technical backgrounds to implement effective disambiguation.

Our simulations also suggest that, depending on the metric, 10~20% disambiguation errors can yield only a few percentage points of measurement error. This raises the possibility that IBD-based studies may not always be as distorted as worried, though the actual number of misidentified entities in past research remains unknown. However, considering that many influential studies applied IBD to all available name strings, effectively corresponding to 100% errors in our simulations, the resulting distortions are likely to be severe. For instance, dramatic increases in average degree, coupled with decreases in unique vertices, may be directly tied to the emergence of power-law degree distributions, where the presence of a few vertices with unrealistically high degrees in the tail is critical. This suggests that power-law fits derived from IBD-based networks warrants closer scrutiny regarding their rigor and validity.

Several limitations remain. First, simulations were run on specific datasets, so the interaction between dataset size and error levels is unclear. Future work should test how resilience of network measures varies under different combinations of data size, merging, and splitting. Second, merging and splitting were tested separately, though merging accounted for most distortions. Future studies should explore their combined effects by varying their proportions and degrees.

## 5 Acknowledgements

This article draws in part from the author's doctoral dissertation [17], with tables and figures adapted from the original work.